\documentstyle[twocolumn,aps,prl]{revtex}

\begin{document}
\draft
\title{Dimer-Type Correlations and Band Crossings in Fibonacci Lattices}
\author{ Ignacio G. Cuesta\cite{paddress} and Indubala I. Satija\cite{email}}
\address{
 Department of Physics, George Mason University,
 Fairfax, VA 22030}
\date{\today}
\maketitle
\begin{abstract}
Fibonacci system with both diagonal
and off-diagonal disorder is mapped to a model with $dimer$
type `defects'. The model exhibit resonance transition
corresponding to zero reflectance condition resulting in extended
states in the system. These
Bloch-type states reside
on the line of inversion symmetry of the energy spectrum
and include states at the band crossings. The resonant states 
at the band crossings
are related to the harmonics of the sine wave and
have transmission coefficient equal to unity in quasiperiodic limit.
This is in contrast to other resonant states
where the transmission coefficient oscillates. 
An exact renormalization scheme confirms the fact that
all resonant states are Bloch type waves
as they are described by trivial attractors of the renormalization flow.

\end{abstract}

\narrowtext

Fibonacci systems have emerged as an interesting class of quasiperiodic
systems exhibiting self-similar fractal wave functions\cite{qp} as well as
Bloch-type extended states.\cite{Wu,macia} In earlier studies\cite{qp}, interest
in these systems was
due to the fact 
that they exhibit cantor spectrum and power-law
decaying wave function and hence are in between periodic and random systems.
Recent interest in the Fibonacci systems is due to the possibiliy of
extended states as was
first discussed briefly by Wu et al\cite{Wu}. It was argued that
the Fibonacci systems with both diagonal and off-diagonal quasiperiodic order
are similar to the 
the binary alloy model which is dual to the dimer model.\cite{dimer} Therefore
Fibonacci system should exhibit the resonance predicted for
the binary alloy model. Recently, Macia et al\cite{macia}
discussed the existence of extended states in these systems by imposing
the condition of commutativity of two non-equivalent
blocks of transfer matrices. They obtained an explicit expression for
the transmission coefficient for a finite size Fibonacci system
and characterized these states as $extended$ $critical$ due to the fact
that in the quasiperiodic limit the states have transmission coefficient ($T$)
which oscillates as the size of the system is varied.
They argued that the quasiperiodic order 
play a key role in determing the extended states
which is in constrast to the analysis of Wu et al where the existence of extended states
is due to a resonance condition unaffected by the long range quasiperiodic
order in the system.

In this paper, the Fibonacci lattices
with both diagonal and off-diagonal disorder are mapped to the lattices
with dimer-type defects. 
Therefore, our results put the Fibonacci systems on the same
footing as the lattices with dimer type correlations\cite{dimer} and hence,
the origin of
extended states in this system is traced to the resonance
transition corresponding to a destructive interference of the
reflected wave between the two neighboring
sites of the dimer defect.
Using this mapping, we derive the resonance condition
and establish the fact that the extended states as discussed by Wu et al are
identical to the propagating states discussed by Macia et al.\cite{Wu,macia}
The fact that the extended states of the Fibonacci system
are in fact Bloch waves on a decimated lattice is
further confirmed by an $exact$ renormalization group (RG) scheme.\cite{KSRG}

Novel result of our studies is the relationship between the resonant states
and the states where the energy bands cross.
We show that the resonant states at the band crossings
are fully transmitting states ($T=1$) in the
quasiperiodic limit and are described by the wave functions that are
related to the harmonics of the sine wave with fundamental Bloch number
equal to the golden mean. Therefore, unlike the rest of states with
oscillating $T$, these states have a special property namely for Fibonacci sizes the perfect transmission limit is well defined.

We consider a tight binding model (TBM) with both diagonal and off-diagonal
disorder, 
\begin{equation}
t_{n+1} \psi_{n+1} + t_{n}\psi_{n-1} + \epsilon_n \psi_n = E \psi_n .
\end{equation}
The Fibonacci system that we study 
is obtained by generating a Fibonacci sequence from two symbols $s$ and $b$ by the substitution
$s \rightarrow b$ and $b \rightarrow bs$.
The corresponding onsite potential takes two values $\epsilon_s$ and $\epsilon_b$.
The resulting lattice can be viewed as made up of $bbs$ and $sbs$ types of
sub blocks.
The off-diagonal couplings are of two types corresponding to
two possible nearest-neighbors and are denoted as $t_{bs}$ and $t_{bb}$.
We choose $t_{bs}=1$ and $t_{bb} = \gamma$ as shown in figure ~1. 

In order to map the Fibonacci system to a dimer model, we decimate
all the $s$ sites
of the lattice with onsite potential $\epsilon_s$. The remaining $b$ type sites
renormalize in two different ways depending upon whether it belongs to
$bbs$ block or $sbs$ block. 
As shown in figure ~1, the resulting 
decimated lattice can be viewed as a pure $r$ type lattice with $d$ type
defects that always appear in pairs. 
The renormalized onsite potential for the regular
and dimer sites
will be denoted as
${\epsilon_r}$ and
${\epsilon_d}$ and the coupling within the
dimers is denoted as $\bar{\gamma}$.
It turns out that the dimers are connected to each other and rest of the
lattice by a coupling of unit strength. The renormalized onsite potential as
well as couplings depend upon $E$ and are related to the corresponding bare values
of the parameters by the following equations.
\begin{eqnarray}
{E_r} &=& (E_b E_s -1)\\
{E_d} &=& (E_b E_s -2)\\
\bar{\gamma} &=& \gamma E_s
\end{eqnarray}
Here $E_x \equiv E-\epsilon_x$, where $x=s, b, r, d$. 

Next we look for a traveling wave solution on the renormalized lattice.
We consider Bloch wave solution where a propagating wave
$e^{2\pi i n k}$ at site $n$ with Bloch number $k$
undergoes phase shifts as it encounters the dimer defects. Let
$\Omega_{rd}$ be the phase shift between the pure lattice sites and the dimers
and $\Omega_{dd}$ be the phase shift  within the dimer. It should be noted
that there is no additional phase shift between two dimers. Substituting these 
traveling wave solutions in the renormalized TBM determines the phase shifts.
 A simple algebra requiring the phase
shifts to be real ( corresponding to nondecaying wave )
determines the resonance condition ,
\begin{eqnarray}
E &=& (\epsilon_b-\gamma^2 \epsilon_s) \over { ( 1-\gamma^2 )}\\
(E-\epsilon_b)(E-\epsilon_s) &= &4 cos^2(k\pi)
\end{eqnarray}

By eliminating $E$, we obtain
\begin{equation}
(\epsilon_b-\epsilon_s) = \frac{2}{\gamma} (1-\gamma^2 ) cos(k\pi)
\end{equation}

The resonance condition we obtain is identical to that of Wu et al\cite{dimer}
for binary alloy.
Furthermore, for $\epsilon_b=-\epsilon_s = \alpha$,
it reduces to the commutativity condition obtained by
Macia et al\cite{macia}. 
Therefore, we conclude that the extended critical states of Macia et
al\cite{macia} are in fact the resonant states.
The renormalization analysis\cite{KSRG} discussed later on in this paper will
provide a confirmation of the fact that the  resonance condition is globally valid.
Therefore, for the parameter values at which the resonance condition is satisfied
we have Bloch wave solutions on the decimated sites. However, on the original
lattice, the Bloch waves exist only on the $b$ sites and
the wave attenuates on $s$ sites and then recovers back on $b$ sites whenever
resonance condition is satisfied.

The resonance criterion shows that 
the off-diagonal disorder, namely $\gamma$ different from unity is crucial
for obtaining extended states in the system. It is interesting to note that
the off-diagonal disorder $\gamma$ also determines the line of
inversion symmetry, $E(n,\alpha)=-E(F_N-n, -\alpha)$ for the spectral plot 
as shown in figure ~2. Here $F_N$ is the size of the
system. This line of inversion symmetry
coincides with the line where the resonance condition is satisfied.
The striking feature of this spectral plot is the existence of band crossings
which form a subset of resonant states.
It turns out that the states at the band crossings corresponds 
a discrete set of energies with Bloch number
$k=n \sigma$ where $n$ is an integer.
Using the expression of the transmission coefficient given by
Macia et al\cite{macia} $T=\frac{1}{ (1+c^2sin (F_N\pi k))}$, 
where $c$ is a function of parameters of the system, we see
that these states corresponds to $T=1$ as the size of the system $F_N$
approaches $\infty$. It should be noted that this approach to quasiperiodic limit
is using Fibonacci sizes only.
Therefore, the states at the band crossings have a well defined 
quasiperiodic limit.
This is in contrast to the rest of the resonant states 
where $T$ oscillates between its minimum value ( corresponding
to $sin(F_N \pi k)=1$ ) and the maximum value equal to
unity, as the size of the system changes. 

It turns out that the resonant states at the band crossings 
are related to the  harmonics of the
sine wave when expressed in terms of a continuous variable 
$\theta=\{ n\sigma + \phi_0 \}$
(where the brackets
denote the fractional part and $\phi_0$ is an arbitrary constant) except
for a discontinuity at $\sigma^2$. 
\begin{eqnarray}
\psi_{n}(\theta) &=&sin(n \theta \pi) , 0 \le \theta < \sigma^2 \\
&=& \frac{1}{\gamma} sin(n \theta \pi) , \sigma^2 \le \theta < 1
\end{eqnarray}
In contrast to $n$-even case, for $n$-odd, the wave function is a double valued function of $\theta$.
This analytical expression was first observed in our numerical
computation of the wave function and is shown in figure ~3. 

The discrete onsite potential
$\epsilon_n$ of the Fibonacci system can be viewed as a two-step function of the variable
$\theta$ with a discontinuity at $\theta=\sigma^2$ where the potential jumps
from -$\alpha$ to $\alpha$ . The states at the band crossings share the discontinuity
of the on-site potential. This characteristics is true for all resonant states
including those that do not correspond to the band crossings.
Such states are typically described by a wave function which is a multivalued
function of $\theta$ with a discontinuity at $\sigma^2$.

The nature of discontinuity in the wave function can be understood from our figure 1.
As we discussed earlier, at resonance, the Bloch wave solutions exist on the decimated
model with $r$ and $d$ type sites after the $s$ sites have been eliminated. It
is precisely
at these $s$ sites that the wave first decays in amplitude by $\gamma$
and then recovers. Therefore, at resonance, although we have Bloch waves
on the decimated model, the original lattice consists of waves that decays
but recovers again. Therefore, for infinite systems these waves are always fully
propagating modes and there is no net attenuation.

We have described two types of resonant states:
the states which are related to all harmonics of the sine functions
and have the transmission coefficient equal to unity 
in quasiperiodic limit
and the states where the transmission coefficient oscillates as we
approach quasiperiodic limit.
We next seek a better characterization of the resonant states  using our
recently developed RG approach. This methodology provides an independent
confirmation of the existence of extended states in the system.
Here the extended states are distinguished from the critical states by
the trivial attractor of the renormalization flow.
This approach  was recently used to demonstrate the existence of
Bloch wave type phonon modes in
the supercritical incommensurate Frenkel-Konterova (FK) model.\cite{FK}

Basic idea underlying the renormalization scheme for the quasiperiodic system
with golden mean incommensurability is
to decimate out
 all lattice sites except
those labeled by the Fibonacci numbers $F_m$. 
Starting from an
arbitrary initial site $n$,
at the $m^{th}$ step, the decimated TBM can be written as
\begin{equation}
\psi(n+F_{m+1})=c_m(n)\psi(n+F_m) + d_m (n) \psi(n). \label{Dec}
\end{equation}
Using the defining property of the Fibonacci numbers,
$F_{m+1} =F_m + F_{m-1}$ with $(F_0 =0, F_1 =1)$, the following recursion relations
are obtained analytically
for the decimation functions $c_m$ and $d_m$,
\begin{eqnarray}
c_{m+1} (n)&=& c_m(n+F_m) c_{m-1} (n+F_m)- \frac{d_{m+1}} {d_m(n)} \\
d_{m+1} (n)&=& -d_m (n)
[d_m (n+F_m )+\nonumber \\
& & c_m(n+F_m) d_{m-1} (n+F_m)] c_m^{-1} (m).
\end{eqnarray}

These recursion relations
can be iterated numerically for a large number of decimation
steps limited only by the precision of the parameters and the energy $E$ and the
machine precision. For resonant states, since the energies are known
analytically, the asymptotic behavior of the RG flow can be obtained
with very high precision.
Our previous studies have shown that the extended
eigenfunctions lead to an asymptotic  trivial attractors of the RG flow
while the critical states at the band edges are characterized by nontrivial
attractors. In the localized phase,
the RG flow decays to zero or becomes unbounded. 
Therefore, by studying the variation in the renormalization
attractor as the parameters $\alpha$ and $\gamma$ are varied,
we can determine the nature of eigenstates of the system.

Figure ~4 shows the renormalization attractor as $\alpha$ varies for
a fixed value of $\gamma$ for states corresponding to the minimum energy 
($k=0$).
There exists a symmetric 4-cycle
which varies continuously as the parameter $\alpha$ changes. This implies that
the wave function repeats itself at every $4^{th}$ Fibonacci site and hence
describes a translational invariance in Fibonacci space for quasiperiodic system.
The variations in the RG attractor implies that the scaling properties of
the critical wave function changes continuously as the parameter $\alpha$ changes.
For a special value of the parameter which satisfy the resonance condition,
this nontrivial RG attractor
degenerates to a trivial cycle $\pm \sigma^{-1}$, determined by the underlying
incommensurability. Therefore, the existence of a $crossings$ in the RG attractors
signal the resonance transitions. Further studies of resonant states
other than those at band crossing shows that all such states are
described by trivial attractors of the renormalization flow. The period of
these attractors vary as the parameters change but the RG flow always settle
asymptotically at values which are $1,0, \infty$ or powers of $\sigma$.
For example, for
for $E=-5/6$, the RG flow converges to $-1,-\sigma,-1,0,\infty,1,1/\sigma,1,0,..$.
It should be noted that in contrast to the resonant states, off-resonant states
are always characterized by a non-trivial RG attractors.
The critical states at the band edges
are characterized by a period-4 cycle while the rest of the critical states
are described by strange attractors of the RG flow. 

In summary, the RG analysis confirms the fact that quasiperiodic Fibonacci systems
exhibit Bloch wave type solutions. Appropriate decimation scheme traces the origin
of these states to hidden dimer type correlations in the system. Therefore all
states that satisfy resonance condition are Bloch waves on a decimated lattice.
These includes the semi transparent states of finite Fibonacci lattices
discussed previously.\cite{macia} For infinite size system, these states
are fully transparent states with oscillating $T$ as there is no net attenuation.
Intriguing result of our paper is the existence of extended states
with Bloch number equal to the multiple of the golden mean. In contrast
to rest of the resonant states, these states which reside at the band crossings
have transmission coefficient equal to unity in the quasiperiodic limit.
Therefore, our studies provide
a clear characterization of extended states in the Fibonacci system and shows
that the existence of fully propagating
states is not related to the quasiperiodic long range order of the system.

The research of I.I.S. is supported by National Science
Foundation Grant No. DMR~097535. 
IIS would like to thank Enrique Macia for fruitful correspondence.
IGC would like to thank for the hospitality
during his visit to the George Mason University.

\begin{figure}
\caption{ Fibonacci lattice with two types of diagonal couplings
defining a lattice with $b$ and $s$ types of atoms. The springs denote the
coupling $\gamma$ while straight lines denote unit coupling. Renormalized lattice
after the $s$ type sites are decimated. The resulting lattice can be viewed as a
regular $r$ type lattice with $d$ type de wh always appear in pairs.
}.
\label{fig1}
\end{figure}

\begin{figure}
\caption{Band spectrum as a function of $\alpha$ for a fixed $\gamma$ equal
to $2$ in (a) and is equal to unity in (b). The resonant states
lie on the dashed line which happens to be the line of inversion symmetry
and passes through the band crossings. In the absence of off-diagonal disorder,
the resonance and band crossings happen at trivial value of $\alpha$.}
\label{fig2}
\end{figure}

\begin{figure}
\caption{Figure shows the wave functions at
four different band crossings as a function of $\theta$:
(a) shows the wave function for $n=2,4$ which is related to the first and
second harmonics of the sine wave, and
(b) corresponds to $n=1,3$. Unlike the even $n$ case, the wave function
is not a single valued function of $\theta$. The wave functions show the discontinuity
of the underlying potential at golden mean.}
\label{fig3}
\end{figure}

\begin{figure}
\caption{ Renormalization attractor as a function of $\alpha$
showing how the period-4 limit cycle
of the decimation function $c_n$ changes as $\alpha$ varies for the states
corresponding to minimum energy($k=0$) at a fixed value of $\gamma=2$.
The existence of resonant
states appears as a crossing as the nontrivial cycle degenerates
to a trivial 2-cycle $\pm \sigma^{-1}$.}
\label{fig4}
\end{figure}

\end{document}